\documentclass[12pt]{iopart}
\usepackage{graphicx}

\begin{document}

\title{Supersymmetric Bragg gratings}

\author{Stefano Longhi }

\address{Dipartimento di Fisica, Politecnico di Milano and Istituto di Fotonica e Nanotecnologie del Consiglio Nazionale delle Ricerche, Piazza L. da Vinci
32, I-20133 Milano, Italy}
\ead{longhi@fisi.polimi.it}
\begin{abstract}
 The supersymmetric (SUSY) structure of coupled-mode equations that describe scattering of optical waves in one-dimensional Bragg gratings is highlighted. This property  can find applications to the synthesis of special Bragg filters and distributed-feedback (DFB) optical cavities. In particular, multiple SUSY (Darboux-Crum) transformations can be used to synthesize DFB filters with any desired number of resonances at target frequencies. As an example, we describe the design of a DFB structure with a set of equally-spaced resonances, i.e. a frequency comb transmission filter.   
\end{abstract}



\maketitle

\section{Introduction}

In field theory, supersymmetry (SUSY) is a model  that relates fermions to bosons in order to obtain a unified description of all basic interactions of nature.
While SUSY finds severe physical and mathematical obstacles as a realistic field theoretical model, SUSY quantum mechanics, i.e. the application of the SUSY superalgebra to quantum mechanics as opposed to quantum field theory, has offered deeper insights into various aspects of standard non-relativistic quantum mechanics, creating new areas of research in quantum mechanics itself \cite{Cooper}. For example, SUSY clarifies why certain one-dimensional potentials are analytically solvable and others are not,  helps in finding new solvable potentials, exemplifies the factorisation (Darboux) method, and generates potentials with similar scattering properties.\par
 Owing to the formal analogy between the Helmholtz equation of monochromatic light waves in dielectric media and the stationary non-relativistic Schr\"{o}dinger equation, the methods of SUSY have been recently transferred into the optical context \cite{S1,S1bis,S2,S3} to synthesize dielectric optical media with tailored scattering and localization properties of potential interest to a wide variety of applications.  SUSY optical structures can be used to realize global phase matching, efficient mode conversion, spatial multiplexing \cite{S1,S4}, as well as to design transparent defects, interfaces and optical intersections \cite{S5,S6,S6bis}. SUSY can be applied to discretized light in waveguide lattices as well \cite{S7,S8}. Periodic optical media, such as distributed
feedback (DFB) structures in  fibers or waveguides, provide an important class of optical devices. While SUSY can be applied in principle to periodic potentials (see, for instance \cite{S5,S6,S8bis}), analytical results in this case are extremely difficult because of the lack of exact analytical forms for the energy dispersion curves and  Bloch functions.  
Moreover, application of SUSY to locally-periodic potentials, in which the local period and potential depth undergo slow changes, is extremely challenging. For locally-periodic shallow gratings, light dynamics can be described with an excellent accuracy by couple-mode theory (CMT), where the slow changes of the grating period and/or grating depth are simply included in a complex scattering potential \cite{CMT1,CMT2}. In CMT, coupling of the two counter-propagating waves in the medium is described by a set of coupled-mode equations (CME), which can  
 be cast in a form analogous to the Dirac equation of a fermionic particle subjected to an external scalar and/or vectorial potential (see, for instance, \cite{L0,L1,L2}). 
 \par   
 In this work we highlight the SUSY structure of CME in DFB structures, and show that it can be suitably exploited to synthesize Bragg grating filters  with a given number of resonances at desired frequencies. This task is accomplished by application of multiple SUSY (Darboux-Crum) transformations \cite{Crum,DC}, which lead to transparent potentials of Kay-Moses type \cite{KM} as a special case. Interestingly, the grating profile is simply determined by the superpotential of the underlying model. As an example, we apply our method to the synthesis of a DFB filter that supports discrete combs of transmission resonances. Application of SUSY to Bragg gratings thus provides an interesting {\it analytical} tool for the synthesis of DFB filters, which is complementary to other analytical synthesis methods based e.g. on the Gelf'and-Levitan-Marchenko inverse scattering \cite{Mar1,Mar2}.

 \section{Coupled-mode equations and supersymmetry}
 
 \subsection{Coupled-mode theory}
 Light propagation in nonuniform grating structures, such as those realized in optical fibers or waveguides, is a well-known problem in the optics of periodic media. For sufficiently shallow gratings that result only in small reflections within a distance of one wavelength, CMT is an excellent approximation to the
exact problem \cite{CMT1,CMT2}. In an effective one-dimensional model and for first-order Bragg scattering, the longitudinal refractive index modulation can be written as $ n(z)=n_0+\Delta n \; h(z) \cos[2 \pi
z / \Lambda + \phi(z)]$, where $n_0$ is the effective mode index in
absence of the grating, $\Delta n \ll n_0$ is a reference value of
the refractive index change of the grating, $\Lambda$ is the nominal grating
period defining the Bragg frequency $\omega_B=\pi c/(\Lambda n_0)$,
$c$ is the speed of light in vacuum, and $h(z)$, $\phi(z)$ describe
the amplitude and phase profiles, respectively, of the grating. Note that, for a non-sinusoidal modulation of the grating, $h(x) \exp[i \phi(z)]$ is defined by the 
first-order Fourier component of the locally-periodic refractive index \cite{CMT2}. Note also that the CMT reproduces with excellent accuracy the results obtained by exact transfer matrix methods for shallow up to moderate gratings ($\Delta n / n_0 < \sim 0.1$) \cite{Leo}, which is the case of corrugated DFB waveguide filters or fiber gratings.   
To
study Bragg scattering of counterpropagating waves at frequencies
close to $\omega_B$, let $E(z,t)=\{ u(z,t) \exp(-i \omega_B t +2 \pi
i z n_0 / \lambda_0)+ v(z,t) \exp(-i \omega_B t -2 \pi i z n_0 /
\lambda_0)+c.c. \}$ be the electric field in the guiding structure, where
$\lambda_0=2 n_0 \Lambda$ is the Bragg wavelength. The envelopes $u$
and $v$ of counterpropagating waves then satisfy the
following CME \cite{CMT2,CMT3}
\begin{eqnarray}
 \left( \frac{\partial}{\partial z} + \frac{1}{v_g} \frac{ \partial}{\partial t}
\right) u & = & iq(z) v \\
 \left( \frac{\partial}{\partial z} - \frac{1}{v_g} \frac{ \partial}{\partial t}
\right) v  & = & -iq^*(z)u
\end{eqnarray}
where 
\begin{equation}
q(z)= \frac{\pi \Delta n } {\lambda_0} h(z) \exp[i \phi(z)]
\end{equation}
 is the complex scattering potential and $v_g \sim c/n_0$ is the group
velocity at the Bragg frequency. 
\subsection{Supersymmetric structure of coupled-mode equations}
It was noticed in previous works \cite{L0,L1,L2} that Eqs.(1) and (2) can be viewed as a Dirac-type equation with scalar and vectorial couplings 
 for the spinor $(u,v)$, under a suitable choice of the Dirac matrices. SUSY and intertwining (Darboux) methods for the Dirac equation have been 
highlighted in several works (see, for instance, \cite{Cooper,D1,D2,D3} and references therein), however their potential use in optics of periodic media has been so far overlooked (with the exception of Ref.\cite{uffa}.)   
To highlight the SUSY structure of CME, it is worth introducing
the dimensionless variables $x=z/Z$ and $\tau=t/T$ , with
characteristic spatial $Z$ and time $T$ scales defined by
\begin{equation}
Z=\frac{\lambda_0}{\pi \Delta n} \; , \; \; T=\frac{Z}{v_g} \simeq \frac{\lambda_0 n_0}{ \pi \Delta n c}.
\end{equation}
and the new envelopes
\begin{equation}
\psi_{1}(z,t)=\frac{u(z,t) - v(z,t)}
{\sqrt 2} \;, \;  \psi_{2}(z,t)=\frac{u(z,t) + v(z,t)}
{\sqrt 2} .
\end{equation}
For monochromatic waves $(u,v) \sim \exp(-i E \tau)$ with a normalized frequency detuning $E=(\omega-\omega_B)T= 2 \pi (\nu- \nu_B)T$ from the Bragg reference frequency $\omega_B= 2 \pi \nu_B$,
 Eqs.(1) and (2) take the form
\begin{eqnarray}
\left( E - m \right) \psi_1 & = i \hat{P}_{-} \psi_2 \\
\left( E + m \right) \psi_2 & = -i \hat{P}_{+} \psi_1. 
\end{eqnarray}
In Eqs.(6) and (7), the operators $\hat{P}_{\pm}$ are defined by
\begin{equation}
\hat{P}_{\pm}= \pm \frac{\partial}{\partial x}+W(x)
\end{equation}
where $m(x)$ and $W(x)$ are given by
\begin{equation}
m(x)=h(x) \cos [\phi(x)] \; ,\;\; W(x)=-h(x) \sin [\phi(x)].
\end{equation}
Here we focus our attention to the special class of grating profiles such that $m(x)$ is a constant (i.e. independent of $x$), for which the SUSY structure of the Dirac equation follows in a rather straightforward way. The $m(x)$=constant case includes the important class of periodic gratings with no chirp, corresponding to $\phi(x)=-\pi/2$ and thus $m(x)=0$ and $W(x)=h(x)$ [see Eq.(9)]. The $m(x)={\rm const} \neq 0$ may describe, for example, the case of chirped gratings that realize the relativistic (Dirac) harmonic oscillator \cite{L1}. Under such an assumption, from Eqs.(6) and (7) it readily follows that $\psi_1$ and $\psi_2$ satisfy two stationary Schr\"{o}dinger equations    
 \begin{equation}
 \hat{H}_1 \psi_1= \epsilon \psi_1 \; , \;\; \hat{H}_2 \psi_2= \epsilon \psi_2
 \end{equation}
 with the same eigenvalue
 \begin{equation}
 \epsilon=E^2-m^2
 \end{equation}
and with partner Hamiltonians $\hat{H}_1$ and $\hat{H}_2$ defined by
\begin{eqnarray}
\hat{H}_1 & = & \hat{P}_- \hat{P}_+ = -\frac{d^2}{dx^2}+V_1(x) \\
\hat{H}_2 & = & \hat{P}_+ \hat{P}_- = -\frac{d^2}{dx^2}+V_2(x) 
\end{eqnarray}
where the partner potentials $V_1(x)$, $V_2(x)$ are related to the superpotential $W(x)$ via the usual relations
\begin{equation}
V_1(x)=W^2(x)-\frac{dW}{dx} \; , \;\; V_2(x)=W^2(x)+\frac{dW}{dx}
\end{equation}
Let us now suppose that the potential $V_1(x)$ is constructed such that the Hamiltonian $\hat{H}_1$ sustains a finite number $N$ of bound states at energies $\epsilon_1$, $\epsilon_2$, $\epsilon_3$, ..., $\epsilon_N$, with $\epsilon_1=0$ and $\epsilon_N> \epsilon_{N-1}> ... > \epsilon_2>\epsilon_1$.  Note that, owing the factorization of $\hat{H}_1$ [Eq.(12)],  the fundamental (lowest-order) bound state $\phi_1(x)$ of $\hat{H}_1$, $ \hat{H}_1 \phi_1= \epsilon_1 \phi_1 =0$, is annihilated by the operator $\hat{P}_+$, i.e.  $\hat{P}_+ \phi_1=0$, and the superpotential $W(x)$ is related to $\phi_1$ by the
 simple relation
\begin{equation}
W(x)=-\frac{(d \phi_1/dx)}{\phi_1}.
\end{equation}
 The Hamiltonian $\hat{H}_2$ is isospectral to $\hat{H}_1$, except for the fact that at $\epsilon=\epsilon_1=0$ $\hat{H}_2$ does not sustain a bound state. Hence, apart from the $\epsilon=\epsilon_1=0$ case, the grating eigenvalue equations (6) and (7) sustain bound states at normalized frequency detunings
$ E_n^{\pm}= \pm \sqrt{\epsilon_n+m^2} \; \; (n=2,3,...,N)$, 
 which follows from Eq.(11). The case $\epsilon=\epsilon_1=0$, corresponding to $E= \pm m$, should be directly investigated by considering the original Eqs.(6) and (7). Since $\hat{P}_+ \phi_1=0$, Eqs.(6) and (7) can be satisfied by taking $\psi_1=\phi_1$, $\psi_2=0$ and $E=m$. Therefore the grating sustains $(2N-1)$ bound states at the frequency detunings
 \begin{eqnarray} 
 E_n^{+} & = & \sqrt{\epsilon_n+m^2}  \; \; (n=1,2,3,4,..,N) \\
 E_n^{-} & = & -\sqrt{\epsilon_n+m^2}  \; \; (n=2,3,..,N)
\end{eqnarray} 
 In particular, for a periodic grating without chirp, corresponding to $ \phi=-\pi/2$ and $m=0$, the bound states occur at the normalized frequency detunings
 \begin{equation}
 E_n=0, \pm \sqrt{\epsilon_2}, \pm \sqrt{\epsilon_3}, ..., \pm \sqrt{\epsilon_N}. 
\end{equation}

 \begin{figure}
\includegraphics[scale=0.4]{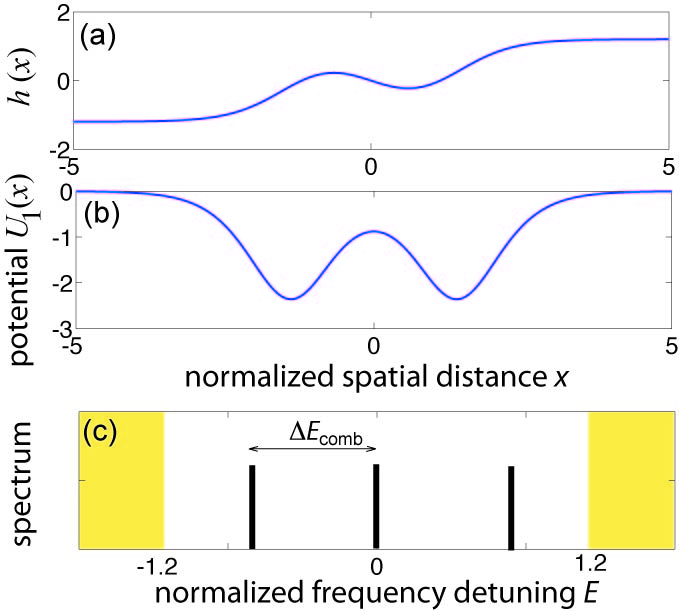}
\caption{(Color online) DFB structure that sustains 3 equally-spaced localized modes as obtained using the Darboux-Crum transformation with $N=2$, $\kappa_1=1.2$ and $\kappa_2=1$. (a) Behavior of the grating amplitude $h(x)$. (b) Behavior of the potential $U_1(x)$. (c) Schematic of the grating spectrum. The three localized modes are equally spaced by $ \Delta E_{comb}=\sqrt{\kappa_1^2-\kappa_2^2} \simeq 0.66$ and are embedded in the stop band of the grating. The shaded areas ($|E|>\kappa_1$) correspond to allowed bands of the grating.}
\end{figure}

 \begin{figure}
\includegraphics[scale=0.4]{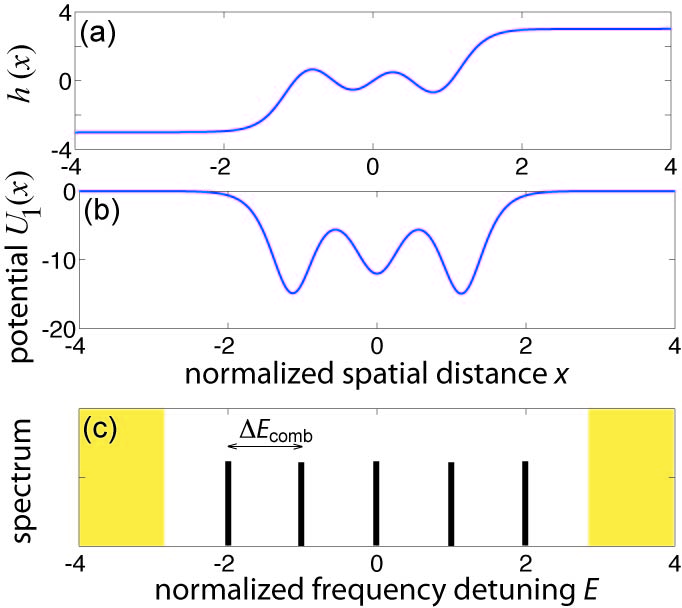}
\caption{(Color online) Same as Fig.1, but for a Darboux-Crum transformation with $N=3$ ($\kappa_1=3$, $\kappa_2=2.8284$ and $\kappa_3=2.2361$). The five localized modes are equally spaced by $ \Delta E_{comb}=\sqrt{\kappa_1^2-\kappa_2^2}=1$.}
\end{figure}

 \section{Darboux-Crum transformations and synthesis of frequency comb DFB filters}
In this section we apply the SUSY structure of grating equations to the synthesis of a DFB structure that realizes a frequency comb transmission filter. For the sake of definiteness, we will consider the case of a periodic grating, i.e. $m(x)=0$ and $W(x)=h(x)$. A powerful and very general method to generate a Hamiltonian $\hat{H}_1$ with a desired number of bound states at target energies is the application of a cascade of SUSY (Darboux) transformations to a reference Hamiltonian, a procedure  known as the Darboux-Crum theorem \cite{Crum,DC} that we briefly review here for the sake of completeness. Let $\hat{H}_0=-\partial^2_x+U_0(x)$ be a reference Hamiltonian and let $\hat{H}_0 \psi(x)= E \psi(x)$, $\hat{H}_0 \varphi_j(x)=\epsilon_j  \varphi_j(x)$ ($j=1,2,...,N$), where $\epsilon_j$ are arbitrary numbers that should not necessarily belong to the point spectrum of $\hat{H}_0$ and $\varphi_j$ are linearly-independent associated functions. The functions $\left\{ \varphi_j(x), \epsilon_j \right\}$ are called {\it seed functions}. Let us then introduce the following functions
\begin{eqnarray}
\xi (x) & = & \frac{\mathcal{W}[\varphi_1,\varphi_2,...,\varphi_N, \psi ](x)}{\mathcal{W}[\varphi_1, \varphi_2,..., \varphi_N](x)} \\
\phi_j (x) & = & \frac{\mathcal{W}[\varphi_1,\varphi_2,..., \tilde{\varphi_j}, ...  \varphi_N, \psi ]}{\mathcal{W}[\varphi_1, \varphi_2,..., \varphi_N]} \;\;\; (j=1,2,...,N)
\end{eqnarray}
where $\mathcal{W}[\psi_1,\psi_2,...,\psi_s]$ is the Wronskian of the  $s$ functions $\psi_1$, $\psi_2$,..., $ \psi_s$, and  $\mathcal{W}[\psi_1,\psi_2,..., \tilde{\psi}_j, ... , \psi_s]$ means that  the function $\psi_j$ should be deleted in the sequence $(\psi_1,\psi_2,..., \psi_s)$. The Darboux-Crum theorem states that 
\begin{equation}
\hat{H}_1 \xi(x)  =  \epsilon \xi (x) \; \; , \; \;  \hat{H}_1 \phi_j(x)  =  \epsilon_j \phi_j (x) \; \; \; (j=1,2,...,N)
\end{equation}
where the new Hamiltonian $\hat{H}_1=-\partial^2_x+U_1(x)$ is defined by the potential
\begin{equation}
U_1(x)=U_0(x)-2 \frac{d^2}{dx^2} {\rm log} \left|  \mathcal{W} [ \varphi_1, \varphi_2, ..., \varphi_n ](x)  \right| .
\end{equation}
To avoid singularities in the potential $U_1(x)$, the Wronskian $\mathcal{W} [ \varphi_1, \varphi_2, ..., \varphi_n ](x)$ should not vanish on the real $x$ axis.
In practice the Darboux-Crum theorem ensures that the two Hamiltonians $\hat{H}_0$ and $\hat{H}_1$, with potentials $U_0(x)$ and $U_1(x)$ related by Eq.(22), are essentially isospectral, apart from the energies $\epsilon_1$, $\epsilon_2$, ..., $\epsilon_N$ that can be added and/or deleted from the spectrum of the original Hamiltonian. For example, if $\epsilon_j$ ($j=1,2,...,N$) do not belong to the point spectrum of $\hat{H}_0$ and the seed functions $\varphi_j(x)$ are unbounded as $x \rightarrow \pm \infty$, then the associated functions $\phi_j(x)$ are normalizable eigenstates of $\hat{H}_1$ and thus the effect of the multiple Darboux transformations is to add the energies $\epsilon_j$ to the point spectrum of the original Hamiltonian. A special case is the one corresponding to a constant potential for the original Hamiltonian, for which application of the Darboux-Crum theorem yields a class of reflectionless potentials of the Kay-Moses kind \cite{KM}.  Specifically, let us assume  $U_0(x)=\kappa_1^2$ and $\epsilon_1=0$, $\epsilon_2=\kappa_1^2-\kappa_2^2$, ..., $\epsilon_N=\kappa_1^2-\kappa_N^2$, where $ \kappa_1 > \kappa_2 > ... > \kappa_N>0$, and let us take $\varphi_j(x)=\exp(\kappa_j x )+(-1)^jc_j \exp(-\kappa_jx)$, with $c_j>0$. The potential $U_1(x)$ of the associated Hamiltonian obtained from the Darboux-Crum theorem then belongs to the class of reflectionless Kay-Moses potentials \cite{DC}. Such potentials can be written rather generally as \cite{KM}
\begin{equation}
U_1(x)=\kappa_1^2-2 \frac{d^2}{dx^2} {\rm log} \left(  {\rm det} \mathcal{A}  \right)
\end{equation}
where the $N \times N$ matrix $\mathcal{A}$ is defined by
\begin{equation}
\mathcal{A}_{n,m}=\delta_{n,m}+\sqrt{A_nA_m} \frac{\exp[i(\kappa_n+\kappa_m)x]}{\kappa_n+\kappa_m}.
\end{equation}
In Eq.(24) $A_n$ are arbitrary real and positive numbers. In particular, with the choice 
\begin{equation}
A_{n}=2 \kappa_n \prod_{m \neq n} \frac{ \kappa_n+\kappa_m}{|\kappa_n-\kappa_m|}
\end{equation}
 the potential $U_1(x)$ turns out to be symmetric, $U_1(-x)=U_1(x)$. The potential (23) sustains $N$ bound states $\phi_j(x)$ at energies $\epsilon_j=\kappa_1^2-\kappa_j^2$ ($j=1,2,...,N$), which can be found from Eq.(20) or as the solutions of the following linear system of equations \cite{KM}
\begin{equation}
\phi_j(x)+A_j \exp( \kappa_jx) \sum_{\nu=1}^{N} \frac{\exp(\kappa_{\nu} x )}{\kappa_j+\kappa_{\nu}} \phi_{\nu}(x)+A_j \exp(i \kappa_j x)=0.
\end{equation}
 The asymptotic behavior of $h(x)=W(x)$ as $ x \rightarrow \pm \infty$ turns out to be given by
 \begin{equation}
 h(x) \sim -\kappa_1 \; \; {\rm as} \; \; x \rightarrow -\infty \;,\;\;\; h(x) \sim \kappa_1 \; \; {\rm as} \; \; x \rightarrow \infty. 
 \end{equation} 
 This means that the Bragg grating is asymptotically a periodic and infinitely-long grating of strength $\kappa_1$, and the frequencies of the bound modes fall within the stop band $(-\kappa_1,\kappa_1)$ of the grating.

\begin{figure}
\includegraphics[scale=0.4]{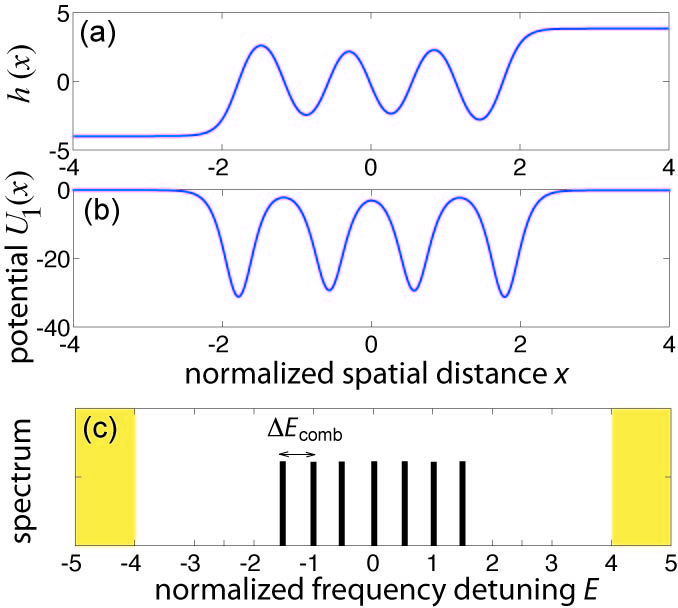}
\caption{(Color online) Same as Fig.1, but for a Darboux-Crum transformation with $N=5$ ($\kappa_1=4$, $\kappa_2=3.9686$, $\kappa_3=3.8730$, and $\kappa_4=   3.7081$). The seven localized modes are equally spaced by $ \Delta E_{comb}=\sqrt{\kappa_1^2-\kappa_2^2}=0.5$.}
\end{figure}

\par
The  Darboux-Crum transformation can provide a useful analytical tool for the synthesis of Bragg gratings with target spectral features. It should noted that, as compared to other analytical methods of grating synthesis based e.g. on the Gelf'and-Levitan-Marchenko inverse scattering method \cite{Mar1,Mar2} or on the similar method of resonance mode expansion \cite{Mar3}, in the Darboux-Crum method the scattering potential $q(x)$ does not  vanish at $x \pm \infty$ [see Eq.(27)], i.e. the grating is synthesized from an underlying uniform and infinitely-extended grating structure by the introduction of defective modes. This is especially useful in the synthesis of DFB cavities that sustain a target number of modes. As an example, we   apply the Darboux-Crum transformation to the synthesis of DFB frequency comb transmission filters. Integrated optical devices such as DFB frequency comb grating filters can be useful for several applications, for example for the minimization of lasing wavelength errors
in DFB lasers, for frequency selection in multiwavelength
excitation and  coupling of lasing and nonlinear
cavities, and for the realization of integrated and spectrally
tailored sources of phase-locked lasing modes (see, for instance,  \cite{APP1,APP2,APP3,APP4} and references therein). To realize an integrated DFB cavity sustaining a comb made of  $(2N-1)$ axial modes with frequency spacing $\Delta \nu_{comb}= \Delta E _{comb}/(2 \pi T)$, let us  consider a periodic index grating (i.e. $\phi(x)= \pi/2$) and let us apply the previously described synthesis method by selecting the sequence of numbers $\kappa_1$, $\kappa_2$, ..., $\kappa_N$ according to
\begin{equation}
\kappa_j=\sqrt{\kappa_1^2- [2 \pi (j-1)T \Delta \nu_{comb}]^2}
\end{equation}
 ($j=1,2,...,N$), where the time scale $T$ is defined by Eq.(4) and where $\kappa_1$ is taken arbitrarily, but larger than $2 \pi (N-1) T  \Delta \nu_{comb}$. The potential $U_1(x)$ and corresponding bound states $\phi_j(x)$ can be obtained from Eqs.(23-26). The grating profile $h(x)=W(x)$ is then retrieved from the expression of the lowest bound state $\phi_1(x)$ using Eq.(15).
 A simple expression for $h(x)$ can be given for the case  $N=2$, i.e. for a DFB cavity sustaining three modes:
 \begin{equation}
 h(x)=\frac{-\kappa_1^2 \sinh(\kappa_1 x)+\kappa_2^2 \sinh(\kappa_1x) {\rm sech}^2(\kappa_2 x) + \kappa_1 \kappa_2 \tanh(\kappa_2 x) \cosh(\kappa_1 x)}{\kappa_2 \tanh (\kappa_2 x) \sinh(\kappa_1 x) -\kappa_1 \cosh(\kappa_1 x)} \;\;\;
 \end{equation}
 the frequency spacing being given by 
 \begin{equation}
 \Delta \nu_{comb}=\frac{\Delta E _{comb}}{2 \pi T}= \frac{\sqrt{\kappa_1^2-\kappa_2^2}}{2 \pi T}.
 \end{equation}
Figures 1, 2 and 3 show typical examples of grating profiles $h(x)$ for DFB cavities that sustain frequency combs made of 3 (Fig.1), 5 (Fig.2) and 7 (Fig.3) modes, as obtained from the Darboux-Crum transformations for $N=2$, $N=3$ and $N=4$, respectively. \par In a DFB structure of finite
length $L$, the ideal amplitude profile $h(x)$ must be truncated after some length far from the 'defective' region, i.e. $h(x)=0$ for
$|x|<L/(2Z)$.
The effect of grating truncation is  that
bound states of the potential $U_1(x)$ become resonance
modes with a finite lifetime (see, for instance, \cite{L1}).
In a passive DFB structure, i.e. in the absence of gain, 
the resonance modes can be observed
as narrow transmission peaks embedded in
the stop band of the grating, at the normalized detunings $E=E_n=(\omega-\omega_B)T$ from the Bragg frequency $\omega_B$ 
according to Eq.(18). This is shown, as an example, in Fig.4(a) for the case of the DFB cavity of Fig.1 sustaining a comb of 3 modes. The figure shows the transmission spectrum $T(E)=|t(E)|^2$ of a periodic DFB structure with the index profile $h(x)$ given in Fig.1(a), taking $h(x)=0$ for $|x|>L/(2Z)=5$. The spectral transmission $t(E)$ versus the normalized frequency detuning $E$ has been numerically computed using a standard transfer matrix method (see, for instance, \cite{CMT3}). In Fig.4(a), the three narrow resonance peaks embedded in the stop band of the grating correspond to the three resonance states of the comb. Owing to grating truncation, other resonance modes appear, however they are high-threshold modes and can be disregarded. The resonance modes of the truncated grating can be numerically computed by looking for the poles of the spectral transmission $t(\tilde{E})$ in the complex $\tilde{E}$ plane, with $\tilde{E}=E-iE_i$ and $E_i>0$ (see, for instance, \cite{L0}). The real part $E$ of the $\tilde{E}$ defines the normalized oscillation frequency detuning of the resonance mode, whereas the imaginary part $E_i$ of $-\tilde{E}$ is the corresponding normalized threshold gain. For a DFB structure with a uniform gain $g$ per unit length, in physical units the threshold gain of each mode is given by
\begin{figure}
\includegraphics[scale=0.4]{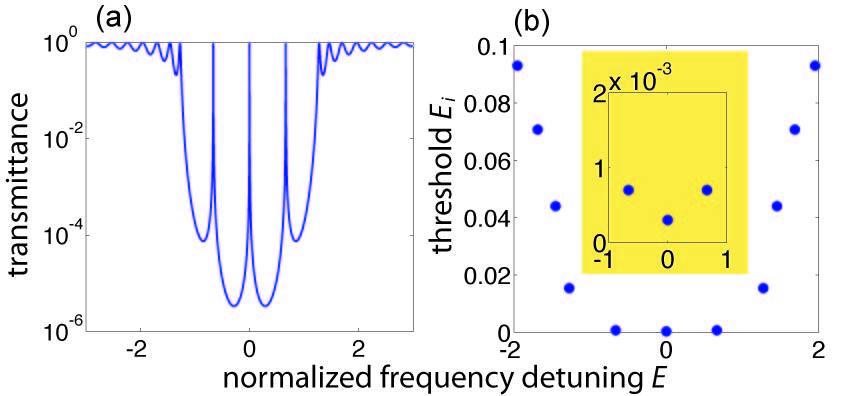}
\caption{(Color online) (a) Numerically-computed transmission spectrum ($|t(E)|^2$) of the DFB filter with index profile shown in Fig.1(a) and with $h(x)=0$ for $|x|>5$. (b) Numerically-computed poles of the spectral transmission $t( \tilde{E})$ in the complex $\tilde{E}=E-iE_i$ plane. The real ($E$) and imaginary ($E_i$) parts the poles define the normalized lasing frequency and lasing threshold of the various DFB cavity modes. The inset in (b) is an enlargement of the three low-threshold poles that correspond to the three bound states of Fig.1(c).}
\end{figure}
\begin{equation}
g_{th}=\frac{E_i}{Z}=\frac{\pi \Delta n E_i}{\lambda_0}
\end{equation}
The numerically-computed poles of $t(\tilde{E})$, i.e. normalized frequency detunings $E$ and normalized thresholds $E_i$ of the various DFB cavity modes, are shown in Fig.4(b). Note that the three comb modes show a very low lasing threshold [$E_i<7 \times 10^{-4}$; see the inset of Fig.4(a)], whereas the other DFB cavity modes have a much higher lasing threshold ($E_i>0.015$, i.e. more than 20 times higher). To get an idea of the design parameters of the DFB structure in real physical units, let us consider a grating realized in a silicon-on-insulator (SOI) waveguide \cite{B1,B2,B3,B4} with Bragg wavelength $\lambda_0=1550$ nm.
Taking $n_0 \sim n_g \sim 3.4$, a first-order grating period $\Lambda \simeq 228$ nm is required \cite{B1}. For a frequency comb spacing $\Delta \nu_{comb}=100$ GHz, the refractive index change $\Delta n$ and the temporal and spatial scales $T$,$Z$, as obtained from Eqs.(4) and (30) with $\kappa_1=1.2$ and $\kappa_2=1$, are given by $\Delta n \simeq 5.13 \times 10^{-3}$, $T \simeq 1.056$ ps and $Z \simeq 93.2 \; \mu$m. The full grating length is thus $L=10 Z \simeq 932 \; \mu$m, the effective coupling constant of the grating is $\kappa_{grating}= \pi \Delta n \kappa_2 / \lambda_0 \simeq 124.65 \; {\rm cm}^{-1}$, and the lasing threshold $g_{th}$ of comb modes is smaller than $0.0751 \; {\rm cm}^{-1}$. Such values are realistic for SOI-based waveguide gratings fabricated
using deep ultraviolet lithography. In particular, a graded-index periodic grating with the target profile as defined in Fig.1(a) could be realized by superposition-apodized or phase-apodized methods \cite{B3}, as well as by duty-cycle engineering \cite{B4}. 
\section{Conclusions}
Bragg scattering in grating structures possesses SUSY properties similar to those of the one-dimensional Dirac equation. Such an interesting property can be useful for the synthesis of DFB filters and cavities with desired scattering and localization properties by application of multiple SUSY (Darboux-Crum) transformations. As an example, we designed DFB structures sustaining with a set of equally-spaced resonances, i.e. a frequency comb transmission filter. Further extensions of SUSY methods to Bragg gratings could be envisaged. For example, the SUSY structure of Dirac equation in the most general case \cite{D3} could be considered in the design of DFB cavities with chirped gratings. Also
SUSY can be applied to underlying non-Hermitian equations, which can be useful for the design of DFB cavities that include gain/loss  gratings \cite{uffa}. Further extensions  include application of second-order (Abraham-Moses) SUSY transformations \cite{DC,Abr}, which might be useful for the design of DFB cavities that sustain bound states embedded in the continuum \cite{kepalle}.


\end{document}